\documentclass{elsart}
\def\mib#1{\hbox{\boldmath $#1$}}

\def\CP{{\mathcal P}}
\def\CT{{\mathcal T}}
\def\CO{{\mathcal O}}

\def\CY{{\mathcal Y}}

\def\lesssim{\begin{array}{c} < \\[-6mm] \sim \\ \end{array}}

\def\bK{\mbox{\boldmath $K$}}

\def\bR{\mbox{\boldmath $R$}}
\def\bS{\mbox{\boldmath $S$}}

\def\bk{\mbox{\boldmath $k$}}
\def\bn{\mbox{\boldmath $n$}}
\def\bp{\mbox{\boldmath $p$}}
\def\bq{\mbox{\boldmath $q$}}
\def\br{\mbox{\boldmath $r$}}

\def\bfell{\mbox{\boldmath $\ell$}}

\def\CO{{\mathcal O}}
\def\CP{{\mathcal P}}

\def\bfsigma{\mbox{\boldmath $\sigma$}}
\def\bftau{\mbox{\boldmath $\tau$}}

\def\aagp{{\alpha \alpha^\prime}}

\def\eq#1{Eq.\,(\ref{#1})}

\begin{document}
\begin{frontmatter}

\title{{\bf Single-Particle Spin-Orbit Strengths
of the Nucleon and Hyperons by} \mib{SU_6} {\bf Quark-Model}}

\author{Y. Fujiwara, M. Kohno$^{*}$, T. Fujita,
C. Nakamoto$^{**}$
and Y. Suzuki$^{***}$
}
\address{
Department of Physics, Kyoto University,
Kyoto 606-8502, Japan \\
$\hbox{}^{*}$Physics Division, Kyushu Dental College,
Kitakyushu 803-8580, Japan \\
$\hbox{}^{**}$Suzuka National College of Technology,
Suzuka 510-0294, Japan \\
$\hbox{}^{***}$Department of Physics, Niigata University,
Niigata 950-2181, Japan}

\maketitle

\begin{abstract}
\indent
The quark-model hyperon-nucleon interaction suggests an
important antisymmetric spin-orbit component.
It is generated from a color analogue of the Fermi-Breit
interaction dominating in the one-gluon exchange process
between quarks.
We discuss the strength $S_B$ of
the single-particle spin-orbit potential,
following the Scheerbaum's prescription.
Using the $SU_6$ quark-model baryon-baryon interaction which was
recently developed by the Kyoto-Niigata group, we calculate $NN$,
$\Lambda N$ and $\Sigma N$ $G$-matrices in symmetric nuclear
matter and apply them to estimate the strength $S_B$.
The ratio of $S_B$ to the nucleon
strength $S_N \sim -40~\hbox{MeV}\cdot\hbox{fm}^5$ is $S_\Lambda/S_N
\sim 1/5$ and $S_\Sigma/S_N \sim 1/2$ in the Born
approximation. The $G$-matrix calculation of the model FSS
modifies $S_\Lambda$ to $S_\Lambda/S_N \sim 1/12$.
For $S_N$ and $S_\Sigma$, the effect of the short-range
correlation is comparatively weak against meson-exchange
potentials with a short-range repulsive core.
The significant reduction of the $\Lambda$ single-particle potential
arises from the combined effect of the antisymmetric $LS$ force,
the flavor-symmetry breaking originating from the strange to up-down
quark-mass difference, as well as the effect
of the short-range correlation.
The density dependence of $S_B$ is also examined.
\end{abstract}

\begin{keyword} $YN$ interaction, quark model,
$G$-matrix, hyperon single-particle potential, spin-orbit interaction
\PACS{13.75.Cs,12.39.Jh,13.75.Ev,24.85.+p}
\end{keyword}
\end{frontmatter}


\section{Introduction}

Though the quantum chromodynamics (QCD) is believed
to be the fundamental theory of the strong interaction,
it is still too difficult to apply the QCD directly
to two-baryon systems. 
At this stage a number of effective models have been
proposed to understand the nucleon-nucleon ($NN$) and
hyperon-nucleon ($YN$) interactions from basic
elements of quarks and gluons \cite{CL79}.
Among them the non-relativistic quark model
has a unique feature that enables us to take
full account of the dynamical motion of the two composite baryons
within a framework of the resonating-group method (RGM) \cite{WS84}.
The model describes confinement with a phenomenological
potential and uses the quark-quark ($qq$) residual interaction
consisting of a color analogue of the Fermi-Breit (FB) interaction.
In the last several years, it was found that such a naive model
does not produce medium- and long-range interactions, but 
can give a realistic description of the $NN$ and $YN$ interactions 
if meson-exchange effects are properly taken into account
in the model.

A simultaneous description of the $NN$ and $YN$ interactions
has recently been achieved by two groups.
One is the $SU_6$ quark model, RGM-F \cite{NA95,FU95},
FSS \cite{FU96a,FU96b,FJ98} and RGM-H \cite{FU96b,FJ98},
by the Kyoto-Niigata group,\footnote{Difference of these three models 
lies only in how to deal with the spin-flavor (-color) factors
of the quark-exchange kernel in EMEP.
In FSS and most of RGM-H these factors are exactly calculated,
while in RGM-F they are approximated to be proportional
to those of the exchange normalization kernel.
RGM-H uses the latter approximation
partly: i.e., solely for the isoscalar-type
scalar-meson ($\epsilon$ and $S^*$) exchanges.}
and the other is the $SU_3$-chiral symmetry
quark model \cite{YU95,ZH97,YA98} by the Beijing-T{\"u}bingen group.
In these models, the spin-flavor $SU_6$ or
chiral-symmetric effective meson-exchange potentials (EMEP)
generated from scalar and pseudo-scalar meson exchanges
between quarks are incorporated.
It was found that the flavor-nonet scalar mesons play
an important role in describing the $NN$ and $YN$ interactions
in a single framework with a unique set of model parameters.
We stress that a simultaneous and
realistic description of the $NN$ and $YN$ interactions
is very important, since the experimental data
for the $YN$ interaction are at present very limited,
and thus one has to rely on the theoretical
consistency of the framework in order to make best use of the rich
experimental information on the $NN$ interaction.

One of the features of the quark-model description
for the $NN$ and $YN$ interactions is that the
antisymmetric $LS$ force ($LS^{(-)}$ force) originating from
the FB spin-orbit interaction is considerably strong
in the strangeness $S=-1$ and
the isospin $I=1/2$ channel \cite{SU84,NA93,MO84}.
Since the signs of the ordinary $LS$ force
and the $LS^{(-)}$ force are opposite
in the $\Lambda N$ interaction,
this strong $LS^{(-)}$ force is vital to produce
very small spin-orbit ($\ell s$) splitting
for the $\Lambda$ single-particle (s.p.) states.
This is consistent with the early experimental observation
that the s.p. spin-orbit term in $40 \geq A \ge 12$ nuclei
is almost zero from the analysis
of the recoilless $(K^-, \pi^-)$ reaction. \cite{BO76}
More recently, preliminary results of the $\gamma$-ray spectroscopy
for $\hbox{}^9_\Lambda\hbox{Be}$ and $\hbox{}^{13}_\Lambda\hbox{C}$   
hypernuclei seem to indicate very small $\ell s$ splitting
in these nuclei. \cite{TA99}
In view of the recent progress of experimental techniques,
a quantitative analysis of s.p. $\ell s$ potentials appears important.
The purpose of this paper is to extend the Scheerbaum's 
formulation \cite{SC76} for the nucleon s.p. $\ell s$ potentials
to hyperons interacting with nucleons via the non-local interaction,
and to examine in detail the s.p. $\ell s$ strengths
of $N$, $\Lambda$ and $\Sigma$, first in the Born approximation
of the quark-exchange kernel,
and then in the $G$-matrix calculation
for our realistic quark-model $NN$ and $YN$ interactions.

Since the spin-orbit interaction between baryons
is essentially short-ranged,
a number of authors have payed attention
to the FB $LS$ force, trying to understand its microscopic
origin from the quark degree of freedom. Here we briefly review
some typical investigations, in which the spin-orbit forces
of the $NN$ and $YN$ interactions are treated
in the $(3q)$-$(3q)$ RGM.
In the WKB-RGM localization techniques of the quark-exchange kernel,
Suzuki and Hecht \cite{SU84} calculated $LS$ potentials,
originating from the symmetric ($sLS$) and antisymmetric ($aLS$)
pieces of the FB interaction.\footnote{Here we follow the notation
given in Eq.\,(5.2) of \protect\cite{FU97}.}
They assumed the same
strange and up-down quark masses and neglected the flavor symmetry
breaking (FSB). This restriction was removed in \cite{NA93}.
After the correction of the sign error of the original paper,
they found that the $sLS$ and $aLS$ spin-orbit terms have same
sign and therefore reinforce each other, giving rise to an
attractive spin-orbit potential in the $\hbox{}^3O$ state
and a repulsive potential in the $\hbox{}^3E$ state for
the $NN$ interaction.\footnote{Ref. \cite{VA95}
cites \cite{SU84} erroneously at this point.}
Morimatsu {\em et al.} \cite{MO84} used only the $sLS$ piece, but
took into account the effect of FSB in a simple approximation.
In these studies, a main interest is naturally the $LS^{(-)}$ force
which involves the simultaneous spin-flip and the flavor exchange
of the hyperon and the nucleon,
a typical feature of the non-identical baryon systems.
The potential concept used in \cite{MO84} is not 
based on the RGM kernel, 
but on the energy surface of the so-called
generator-coordinate method (GCM) kernel.
Using the folding procedure for the GCM kernel,
they calculated, for the first time, the quark-model
predictions for the s.p. $\ell s$ potentials
of the nucleon and hyperons in symmetric nuclear matter.
Although their absolute values of the s.p. $\ell s$ strengths
are somewhat too large, they obtained the relative ratio,
$U_N : U_\Lambda : U_\Sigma=1 : 0.21 : 0.55$, which is very close
to our prediction $1 : 1/5 : 1/2$ in the Born approximation
given in this paper.
On the other hand, He, Wang and Wong \cite{HE86} compared the
quark-model potentials with the Paris (for $NN$) and
the Nijmegen potentials in the form of the Born amplitudes.
They explicitly introduced a core radius $c$,
in order to take into account the effect
of the short-range correlation in the Nijmegen hard-core
model D \cite{HCD} and model F \cite{HCF}.
This procedure was also adopted by the J{\"u}lich group
to show the relative strength of the $LS$ and $LS^{(-)}$ forces
in their one-boson exchange potential (OBEP) model \cite{RE94}.
Through all of these studies, it is now generally recognized
that the quark-exchange kernel from the FB interaction
leads to the spin and flavor dependence which is
qualitatively very similar to that of OBEP,
thus yielding a possible alternative to the meson-exchange 
description of the $LS$ interaction by vector and scalar mesons.
\footnote{There exists, however, appreciable quantitative
difference between predictions by various versions
of OBEP and the quark model.
For example, the Nijmegen potentials generally predict a rather
small $LS^{(-)}$ force, compared with that of the quark model.}

Strictly speaking, the $LS$ force cannot be discussed independently
of other pieces of interaction. Apparently the $LS$ force 
is influenced by the description of the short-range correlation
which is different between the meson-exchange model
and the quark model. 
The $LS$ force also depends on
how to derive the s.p. $\ell s$ potentials  
in the finite nuclei from the original $NN$ and $YN$ interactions
in the free space. In fact, the first issue is the major motivation
for any realistic quark models for the $NN$ and $YN$ interactions.
For example, Yang {\em et al.} \cite{YA98} discussed
the difference of the one-gluon exchange (OGE) process
and the scalar-meson nonet exchange (OSE) introduced between quarks
in the framework of the chiral $SU_3$ quark model.
Since their $LS$ force is too weak in the $NN$ sector
because of several reasons, they reinforced the $sLS$ term
of OGE by a factor of 3.1 and that of OSE by a factor of 4.8.  
Through this prescription, they argued that a sizable OGE component,
which would definitely result in a quite strong $LS^{(-)}$ force
in the $I=1/2$ channel, is not favorable, since it leads to
an unphysical resonance in the $\Lambda N$ channel.
This is more or less a correct statement
as long as the $LS$ components of the FB interaction
is concerned. However, our result in the model RGM-H \cite{FU96b,FJ98}
implies that there exists a solution which reproduces
the necessary $LS$ force in the $NN$ interaction without
introducing any enhancement factor,
and still reproduces the observed $\Lambda p$,
$\Sigma^+ p$ and $\Sigma^- p$ differential
cross sections reasonably well.
The main difference between the two models lies
in the choice of the harmonic oscillator
constant $b$ of the $(3q)$ clusters and the magnitude
of the quark-gluon coupling
constant $\alpha_S$.\footnote{The model RGM-H
(nor the other versions, RGM-F and FSS) does not
include the $LS$ component from the scalar-meson exchange,
while it is included in the chiral $SU_3$ quark model.
However, the incorporation of even more sophisticated EMEP
involving vector mesons does not change this situation.
A new version of our quark model
in this direction will be published elsewhere.}
The Beijing, T{\"u}bingen and Salamanca groups
use $b \sim 0.5~\hbox{fm}^{-1}$ and $\alpha_S \sim 0.5$,
while RGM-F, FSS and RGM-H use $b \sim 0.6
~\hbox{fm}^{-1}$ and $\alpha_S \sim 2$.
Since the $LS$ force is short-ranged, it is very sensitive to
the magnitude of the size parameter $b$.
It is sometimes claimed that
our $\alpha_S$ is too big, compared with the QCD coupling
constant, and is contradictory to the experimental fact that
there seems no spin-orbit splitting existing
in the negative-parity excited states of baryons (especially,
the nucleon and $\Delta$). We should, however, keep in mind that
our $\alpha_S$ is merely a model parameter
in the nonperturbed region, which has very little
to do with the real QCD coupling constant. The explicit value is
determined from the best fit to the experimental data in the
present framework.
The second point is the so-called ``missing $LS$ force'' problem
in the $P$-wave baryons.
Fujiwara \cite{FU93} has shown that the seemingly 
small spin-orbit splitting of the $P$-wave baryon spectrum can be
explained by the dispersive effect due to the resonance nature
of these $P$-wave baryons embedded in the baryon-meson continua.
In other words, the missing $LS$ force problem of the $P$-wave
baryons does not necessarily indicate that the FB interaction 
is inappropriate as a residual interaction
of the non-relativistic quark model.

In this paper, we apply the Scheerbaum's discussion \cite{SC76} for
the strength of the nucleon s.p. $\ell s$ potential to the
formulation of the quark-model invariant amplitudes
developed in \cite{FU97}.
The strength factor $S_B$ for the hyperon s.p. $\ell s$ potential
in the Thomas form is explicitly derived.
Two different kinds of approaches
are attempted in the Born approximation.
One is to use the Wigner transform at $\bp=0$ in the WKB-RGM
formalism as an effective local potential in low-energy processes,
and the other is the $P$-wave approximation
for the dominant contribution to the $LS$ invariant amplitudes.
Both methods involve some kind of averaging procedure
for the spatial integrals and leave one momentum as an input parameter.
This momentum dependence, however, is generally very weak.
One can thus adopt the zero-momentum limit,
in which these two methods give the same result,
yielding very simple expressions for $S_B$.
We consider spin-saturated (s.s.) finite nuclei
or symmetric nuclear matter.
The most reliable description of $S_B$ is
therefore formulated through the nuclear-matter approximation
of the $G$-matrix invariant amplitudes.

We present in Section 2 basic formulae to calculate $S_B$.
After introducing two kinds of approximations
to the spatial integrals for the $LS$ Born amplitudes
in Subsection 2.2, a method of $G$-matrix calculation
is discussed in Subsection 2.3.
In Section 3 we give analytic expressions of $S_B$ 
in the simplest Born approximation,
and use them to examine the characteristic structure
of the s.p. $\ell s$ potentials.
The $G$-matrices calculated in nuclear matter are used to
obtain a more realistic estimate for $S_B$.
The strength $S_\Lambda$ turns out to be very small because of the
cancellation between $LS$ and $LS^{(-)}$ components.
The short-range correlation is found to further
reduce $S_\Lambda$ to be less than $(1/10)S_N$.
Section 4 is devoted to a summary.

\vspace{-3mm}

\section{Formulation}

\vspace{-3mm}

\subsection{Strengths of hyperon single-particle spin-orbit potentials}

\vspace{-3mm}

We start from the RGM equation
for the $(3q)$-$(3q)$ system \cite{NA95,FU96b}:
\begin{equation}
\left[~\varepsilon_\alpha + {\hbar^2 \over 2\mu_\alpha}
\left({\partial \over \partial \bR} \right)^2~\right]
\chi_\alpha(\bR)=\sum_{\alpha^\prime} \int d \bR^\prime
~G_{\alpha \alpha^\prime}(\bR, \bR^\prime; E)
~\chi_{\alpha^\prime}(\bR^\prime)\ \ ,
\label{fm1}
\end{equation}
where the $G_{\alpha \alpha^\prime}(\bR, \bR^\prime; E)$ is
composed of various pieces of the interaction kernels
as well as the direct potentials of EMEP.
The subscript $\alpha$ stands for a set of quantum
numbers of the channel wave function; $\alpha=
\left[1/2(11)\,a_1, 1/2(11)a_2 \right]$ $SS_zYII_z; \CP$,
where $1/2(11) a$ is the spin and $SU_3$ quantum number
in the Elliott notation $(\lambda \mu)$, $a=YI$ the flavor label
of the octet baryons ($N=1(1/2),~\Lambda=00,
~\Sigma=01$ and $\Xi=-1(1/2)$),
and $\CP$ is the flavor-exchange phase.
In the $NN$ system with $a_1a_2=NN$, $\CP$ is actually redundant
since $\CP=(-1)^{1-I}$.
The relative energy $\varepsilon_\alpha$ in the
channel $\alpha$ is related to the total energy $E$ of the system
through $\varepsilon_\alpha=E-E^{int}_{a_1}-E^{int}_{a_2}$.
According to \cite{FU97}, we introduce the basic Born
kernel of \eq{fm1} through
\begin{eqnarray}
M_\aagp (\bq_f, \bq_i; E)
& = & \langle\,e^{i \bq_f \cdot \bR}\,\vert
\,G_\aagp (\bR, \bR^\prime; E)
\,\vert\,e^{i \bq_{\,i}\cdot {\bR}^\prime} \rangle \nonumber \\
& = & \langle\,e^{i \bq_f \cdot \bR} \eta_\alpha^{SF}\,\vert
\,G(\bR, \bR^\prime; E)\,\vert\,e^{i \bq_{\,i}\cdot
{\bR}^\prime} \eta_{\alpha^\prime}^{SF} \rangle\ \ ,
\label{fm2}
\end{eqnarray}
%
%
where $\eta_\alpha^{SF}$ is the spin-flavor wave function
at the baryon level, defined in Eq.\,(2.9) of \cite{FU97}.

In the following we restrict ourselves
to the spin-saturated (s.s.) nuclei
and apply the Scheerbaum's prescription
for the s.p. spin-orbit strengths,
first to the quark-exchange
kernel $G_{\alpha \alpha^\prime}(\bR, \bR^\prime; E)$,
secondly to the $G$ matrices obtained by solving the corresponding
Bethe-Goldstone equation \cite{KO99}.
We call the first prescription the Born approximation,
and the second one a realistic calculation.
Suppose $G$ is the quark-exchange
kernel $G(\bR, \bR^\prime; E)$ or the $G$-matrix.
We calculate s.p. energy
\begin{equation}
E^{s.s.}_v = \sum_c~\langle\,vc\,|\,G\,|\,vc-cv\,\rangle\ \ ,
\label{fm3}
\end{equation}
for the spin-orbit interaction.
The two-particle interaction $G$ is assumed
to be expressed as
\begin{eqnarray}
\langle\,\bk_1 \bk_2\,|\,G\,|\,\bk_1^\prime \bk_2^\prime\,\rangle
& = & \delta(\bK_{12}-\bK_{12}^\prime )~\langle\,\bk_{12}\,|
\,G\,|\,\bk_{12}^\prime\,\rangle 
\nonumber \\
& = & \delta(\bK_{12}-\bK_{12}^\prime )~{1 \over (2\pi)^3}
~M(\bk_{12}, \bk_{12}^\prime )
\ \ ,
\label{fm4}
\end{eqnarray}
where $\bK_{12}=\bk_1+\bk_2$ and $\bk_{12}
=(\xi \bk_1-\bk_2)/(1+\xi)$ with $\xi=(M_2/M_1)$ are
the center-of-mass and relative momenta, respectively.
In the case of the $G$-matrix, $M(\bk_{12},
\bk_{12}^\prime )$ may depend on $(\bK_{12})^2$ and
the starting energy as well. 
It is convenient to use the invariant kernel $M^\Omega (\bk_{12},
\bk_{12}^\prime)$, by which the Born kernel \eq{fm2} is
expressed as
\begin{equation}
M(\bk_{12}, \bk_{12}^\prime) = \sum_\Omega
M^\Omega (\bk_{12}, \bk_{12}^\prime)~\CO^\Omega
(\bk_{12}, \bk_{12}^\prime)
\ \ .
\label{fm5}
\end{equation} 
Here we only consider $\Omega=LS,~LS^{(-)}$ and $LS^{(-)}\sigma$
components \cite{FU97}, which are 
represented by the Pauli-spinor invariants
\begin{eqnarray}
& & \CO^{LS} = i \bn \cdot \bS\ ,\quad
\CO^{LS^{(-)}} = i \bn \cdot \bS^{(-)}\ ,\quad 
\CO^{LS^{(-)}\sigma} = i \bn \cdot \bS^{(-)}\,P_\sigma
\ \ ,\nonumber \\
& & \hbox{with} \qquad \bn=[ \bk_{12}^\prime \times \bk_{12} ]\ ,\quad 
\bS={1 \over 2}(\bfsigma_1+\bfsigma_2)\ ,\quad
\bS^{(-)}={1 \over 2}(\bfsigma_1-\bfsigma_2)
\ \ ,\nonumber \\
& & \hbox{and} \qquad
P_\sigma={1+\bfsigma_1 \cdot \bfsigma_2 \over 2}\ \ .
\label{fm6}
\end{eqnarray} 
The invariant kernel $M^\Omega (\bk_{12},\bk_{12}^\prime)$ in
\eq{fm5} consists of various types of spin-flavor
factors $X_{\CT}^\Omega$ and
the spatial functions $f_{\CT}^\Omega (\theta)$ calculated 
for the quark-exchange kernel of the FB interaction.
These are explicitly given in \cite{FU97} and
Appendix A.
When the contribution from the exchange Feynman diagram
is incorporated with the exchange operator $P_\sigma P_F P_r$,
the total Born kernel is expanded as
\begin{eqnarray}
& & M(\bk_{12}, \bk_{12}^\prime)
- M(\bk_{12}, -\bk_{12}^\prime)\,P_\sigma\,P_F\ \ \nonumber \\
& & = \sum_\Omega
M^{\Omega~\hbox{total}} (\bk_{12}, \bk_{12}^\prime)
~\CO^\Omega(\bk_{12}, \bk_{12}^\prime)
\ \ ,
\label{fm7}
\end{eqnarray} 
with the matrix element in the isospin basis
\begin{eqnarray}
& & \langle\,[BN]_{II_z}\,|\,M(\bk_{12}, \bk_{12}^\prime)
- M(\bk_{12}, -\bk_{12}^\prime)\,P_\sigma\,P_F\,|
\,[BN]_{II_z}\,\rangle\ \ \nonumber \\
& & = \sum_\Omega
M_{BB}^{\Omega~\hbox{total}} (\bk_{12}, \bk_{12}^\prime)
~\CO^\Omega(\bk_{12}, \bk_{12}^\prime)
\ \ .
\label{fm8}
\end{eqnarray} 
Here the $LS$ components $M_{BB}^{\Omega~\hbox{total}}
(\bk_{12}, \bk_{12}^\prime)$ for the $NN$ and $YN$ systems are
explicitly given by
\begin{eqnarray}
M_{NN}^{LS~\hbox{total}}(\bk_{12}, \bk_{12}^\prime)
& = & \sum_{\CT}(X_{\CT}^{LS})_{NN} \left[~f_{\CT}^{LS}(\theta)
-(-1)^I\,f_{\CT}^{LS}(\pi-\theta)~\right]\ \ ,\nonumber \\
M_{YY}^{LS~\hbox{total}}(\bk_{12}, \bk_{12}^\prime)
& = & \sum_{\CT} \left[ (X_{\CT}^{LS})_{YY}^{ud}\,f_{\CT}^{LS}(\theta)
+(X_{\CT}^{LS})_{YY}^s\,f_{\CT}^{LS}(\pi-\theta)~\right]\ \ ,\nonumber \\
M_{YY}^{LS^{(-)}~\hbox{total}}(\bk_{12}, \bk_{12}^\prime)
& = & \sum_{\CT} \left[ (X_{\CT}^{LS^{(-)}})_{YY}^{ud}
\,f_{\CT}^{LS}(\theta) \right. \nonumber \\
& & \left. +(X_{\CT}^{LS^{(-)}\sigma})_{YY}^s
\,f_{\CT}^{LS}(\pi-\theta)~\right]\ \ ,\nonumber \\
M_{YY}^{LS^{(-)}\sigma~\hbox{total}}(\bk_{12}, \bk_{12}^\prime)
& = & \sum_{\CT} \left[ (X_{\CT}^{LS^{(-)}\sigma})_{YY}^{ud}
\,f_{\CT}^{LS}(\theta) \right. \nonumber \\
& & \left. +(X_{\CT}^{LS^{(-)}})_{YY}^s
\,f_{\CT}^{LS}(\pi-\theta)~\right] \ \ . 
\label{fm9}
\end{eqnarray} 
We should note that the spin-flavor factors depend on isospin
and the $LS$ function $f_{\CT}^\Omega (\theta)$ given
in \eq{a3} is a function of $\bk_{12}^2$ and $(\bk_{12}^\prime)^2$,
in addition to the relative angle: $\cos \theta=\widehat{\bk}_{12}
\cdot \widehat{\bk^\prime}_{12}$. The sum over $\CT$ in \eq{fm9} is
with respect to the quark-exchange interaction types $\CT=S,~S^\prime$
and $D_+,~D_-$ \cite{KI94}, where the former two come
from the $aLS$ term of the FB interaction and the latter two
from the $sLS$ term (see Eq.\,(5.2) of \cite{FU97}).
For the $YN$ system, the exchange
term (i.e., $f_\CT^{LS}(\pi- \theta)$ term) in \eq{fm9} originates
from the strangeness exchange process and the spin-flavor factors
of the $LS^{(-)}$ and $LS^{(-)}\sigma$ types are interchanged
between the $LS^{(-)}$ and $LS^{(-)}\sigma$ terms.
The s.p. wave functions are expressed as
\begin{eqnarray}   
& & \psi(\bk, s)={1 \over (2\pi)^3} \int\,d\,\bk~e^{-i\bk \br}
\psi(\br, s)\ \ ,\nonumber \\
& & \psi(\br, s)=\sum_{m_\ell m_s} \langle\,\ell m_\ell
\textstyle{1\over 2}m_s |
j m\,\rangle~\phi_{n\ell m_\ell}(\br)\,\chi_{{1 \over 2}
m_s}(s)\ \ ,\nonumber \\
& & \phi_{n\ell m_\ell}(\br)=R_{n\ell}(r)\,Y_{\ell
m_\ell}(\widehat{\br})\ \ ,
\label{fm10}
\end{eqnarray} 
for the valence particle and the core nucleons; $v,~c=n\ell j$.
By noting 
\begin{equation}
\langle\,\bk_1 \bk_2~[BN]^{II_z}\,|\,vc\,\rangle
=\psi_v(\bk_1, s_1)\,\psi_c(\bk_2, s_2)\,\langle~I_v\,I_{vz}
\textstyle{1\over 2}~\tau\,|\,II_z\,\rangle\ \ ,
\label{fm11}
\end{equation} 
with $\tau=1/2$ for $c=p$ and $\tau=-1/2$ for $c=n$ and
taking a sum over $c$ for the core protons ($c_\tau=c_{1/2}$)
and neutrons ($c_\tau=c_{-1/2}$) separately,
\eq{fm3} becomes $E_v^{s.s.}=\sum_\tau  E_{v; \tau}^{s.s.}$ with
\begin{eqnarray}
E_{v; \tau}^{s.s.} & = & \sum_{c_\tau} \sum_I C_\tau^I(B)
~{1 \over (2\pi)^3}
\sum_{\bk_1, \bk_2, \bk_1^\prime, \bk_2^\prime}
\psi_v^\dagger (\bk_1, s_1)\,\psi_{c_\tau}^\dagger (\bk_2, s_2)
~\delta(\bK_{12}-\bK_{12}^\prime)
\nonumber \\
& & \times \sum_\Omega
M_{BB}^{\Omega~\hbox{total}}(\bk_{12}, \bk_{12}^\prime)
\,\CO^\Omega(\bk_{12}, \bk_{12}^\prime)\,
\psi_v (\bk_1^\prime, s_1)
\,\psi_{c_\tau} (\bk_2^\prime, s_2)\ \ .
\label{fm12}
\end{eqnarray} 
The isospin factor defined by 
\begin{equation}
C_\tau^I(B)=\sum_{I_z}\,\langle I_v\,I_{vz}
\textstyle{1\over 2} \tau\,|\,II_z\,\rangle^2 \ \ 
\label{fm13}
\end{equation} 
is given in \eq{a1}.
The implicit spin sum over $s_2$ is easily carried out:
\begin{eqnarray}   
& & \sum_{c_\tau} \psi_{c_\tau}^\dagger (\bk_2, s_2)
\left\{ \begin{array}{c} \bS \\ \bS^{(-)} \\ \end{array} \right\}
\psi_{c_\tau}(\bk_2^\prime, s_2)
={1 \over 2}~\rho_\tau(\bk_2, \bk_2^\prime)
~\bfsigma_1 \ \ ,\nonumber \\
& & \sum_{c_\tau} \psi_{c_\tau}^\dagger (\bk_2, s_2)
~\bS^{(-)} P_\sigma~\psi_v(\bk_1^\prime, s_1)
~\psi_{c_\tau}(\bk_2^\prime, s_2)=0\ \ ,
\label{fm14}
\end{eqnarray}
where the core-density of the protons or neutrons are defined by
\begin{equation}
\rho_\tau(\bk_2, \bk_2^\prime)=2 \sum_{(n\ell m_\ell)\,\in\,c_\tau}
\phi_{n\ell m_\ell}^*(\bk_2)~\phi_{n\ell m_\ell}(\bk_2^\prime)\ \ .
\label{fm15}
\end{equation} 
We find that the $LS^{(-)}\sigma$ term does not contribute
due to the spin-averaging.
After all, we have obtained
\begin{eqnarray}   
& & E_{v; \tau}^{s.s.}=\sum_I\,C_\tau^I(B)
~{1 \over 2(2\pi)^3} \sum_{\bk_1, \bk_2, \bk_1^\prime, \bk_2^\prime}
\psi_v^\dagger (\bk_1, s_1)~\delta(\bK_{12}-\bK_{12}^\prime)
~\rho_\tau(\bk_2, \bk_2^\prime)\ \ \nonumber \\
& & \times \left[
~M_{BB}^{LS~\hbox{total}}(\bk_{12}, \bk_{12}^\prime)
+M_{BB}^{LS^{(-)}~\hbox{total}}(\bk_{12}, \bk_{12}^\prime)
~\right]\nonumber \\
& & \times i\,[ \bk_{12}^\prime \times \bk_{12} ]\cdot \bfsigma_1
~\psi_v (\bk_1^\prime, s_1)\ \ .
\label{fm16}
\end{eqnarray} 
So far we have made no approximation.

We can eliminate the $\bk_2^\prime$ sum in \eq{fm16} through
$\bk_1+\bk_2=\bk_1^\prime+\bk_2^\prime$; i.e., $\bk_2^\prime
=\bk_1+\bk_2-\bk_1^\prime$. If we use the momenta $\bq$ and
$\bp$ defined by
\begin{eqnarray}   
\bq & = & \bk_{12}-\bk_{12}^\prime
=\bk_1-\bk_1^\prime\ \ ,\nonumber \\
\bp & = & \bk_{12}^\prime+{1 \over \xi}\bk_{12}=\bk_1^\prime
-{ 1 \over \xi} \bk_2\ \ ,\nonumber \\
\label{fm17}
\end{eqnarray} 
the outer product $[ \bk_{12}^\prime \times \bk_{12}]$ in \eq{fm16}
can be expressed as
\begin{equation}
[ \bk_{12}^\prime \times \bk_{12}]=-{\xi \over 1+\xi}\left\{
~[\bk_1 \times \bk_1^\prime]+{1 \over \xi}[\bk_2 \times \bq]
~\right\}\ \ .
\label{fm18}
\end{equation} 
The essential point of the Scheerbaum's discussion \cite{SC76} is that
his space integrals $D(q)/q$ and $E(p)/p$ are very smooth functions
with respect to the small values of momentum transfers $q=|\bq|$
and $p=|\bp|$. From this observation, he replaced the integral by 
a constant value $\langle D(q)/q+E(p)/p \rangle$ evaluated at 
an appropriate averaged value $p=q={\bar q}$ and carried out 
the summation over $\bk_1,~\bk_2$ and $\bk_1^\prime$ in \eq{fm16}.
This behavior of the integral is related to
the short-range character of the $LS$ interaction and
the assumption of the locality of the effective spin-orbit potential
adopted there. Though the non-locality of 
the exchange kernel in the present case makes it difficult to 
follow his argument directly, we can make use of the
short-range character of the $LS$ force and assume that the
amplitudes $M_{BB}^{\Omega~\hbox{total}}(\bk_{12},
\bk_{12}^\prime)$ in \eq{fm16} have a very weak $\bk_1,~\bk_2$
and $\bk_1^\prime$ dependence.
As we will see later, this assumption turns
out to be fairly good even in our case. 
Following the same procedure as developed by Scheerbaum,
we can eventually arrive at the s.p. spin-orbit 
potential of the Thomas type:
\begin{eqnarray}   
& & E_{v; \tau}^{s.s.}=\int d\,\br_1 
\,\psi_v^\dagger (\br_1, s_1)~U_\tau(\br_1)
~\psi_v (\br_1, s_1) \ \ ,\nonumber \\
& & U_\tau(\br)=K_\tau~{1\over r}{d \rho_\tau(r) \over d r}
~\bfell \cdot \bfsigma_1\ \ ,
\label{fm19}
\end{eqnarray} 
where the proton ($\tau=1/2$) and neutron ($\tau=-1/2$) densities
are defined by
\begin{equation}
\rho_\tau(r)=2 \sum_{(n\ell)\,\in\,c_\tau}
{4\pi \over 2\ell+1} [\,R_{n\ell}(r)\,]^2\ \ ,
\label{fm20}
\end{equation} 
and the strength factor is given by
\begin{eqnarray}
K_\tau & = & -{1 \over 2}~{\xi \over 1+\xi}
\sum_I C_\tau^I(B) \nonumber \\
& & \times
\overline{\left[\,M_{BB}^{LS~\hbox{total}}(\bk_{12}, \bk_{12}^\prime)
+M_{BB}^{LS^{(-)}~\hbox{total}}(\bk_{12}, \bk_{12}^\prime)
\,\right]}\ \ .
\label{fm21}
\end{eqnarray} 
For the s.s. nuclei with equal proton and neutron
numbers (i.e., $Z=N$), the formulae
in Eqs.\,(\ref{fm19}) and (\ref{fm21}) are further simplified into
\begin{eqnarray} 
U(\br) & = & -{\pi \over 2}~S_B~{ 1\over r}{d \rho(r) \over d r}
~\bfell \cdot \bfsigma\ \ ,\nonumber \\
S_B & = & {1 \over 2\pi}~{\xi \over 1+\xi} 
\sum_I {2I+1 \over 2I_B+1}\nonumber \\
& & \times \overline{\left[\,M_{BB}^{LS~\hbox{total}}
(\bk_{12}, \bk_{12}^\prime)
+M_{BB}^{LS^{(-)}~\hbox{total}}(\bk_{12}, \bk_{12}^\prime)
\,\right]}\ \ ,
\label{fm22}
\end{eqnarray} 
where $\rho(r)=\rho_n(r)+\rho_p(r)$ with $\rho_n(r)=\rho_p(r)$ is
the total density and
the sum formula (\ref{a2}) for $C_\tau^I(B)$  is used.
We call $S_B$ in \eq{fm22} the Scheerbaum factor.

%

\bigskip
\begin{table}[t]
\caption{The spin-flavor factors $X_\CT^B$ as a function
of $\lambda=(m_s/m_{ud})$. Note that $X_{S^\prime}^B=X_S^B$.
The $X_\CT^B$ values for $\lambda=1$ are given in the second line.
The last row implies off-diagonal factors
for the $\Lambda N$-$\Sigma N$ coupling.}
\label{table1}
\bigskip
%
\begin{center}
\renewcommand{\arraystretch}{1.4}
\setlength{\tabcolsep}{2mm}
\begin{tabular}{cccc}
\hline
$B$ & $X_{D_-}^B$ & $X_{D_+}^B$ &  $X_S^B$ \\
\hline
$N$ & ${14 \over 9}$ & $-{10 \over 27}$ & ${16 \over 81}$ \\
\hline
$\Lambda$ & ${2 \over 9\lambda}\left(2+{1 \over \lambda}\right)$
& $-{1 \over 9\lambda}\left(2+{1 \over \lambda}\right)$
& ${1 \over 18\lambda}\left(2-{1 \over \lambda}\right)$ \\
$\lambda=1$ & ${2 \over 3}$ & $-{1 \over 3}$ & ${1 \over 18}$ \\
\hline
$\Sigma$ & ${2 \over 3\cdot 81}\left(106-{6 \over \lambda}
-{1 \over \lambda^2}\right)$
& $-{1 \over 81}\left(18 -{10 \over \lambda}
-{3 \over \lambda^2}\right)$
& ${1 \over 6\cdot 81}\left(26+{42 \over \lambda}
-{7 \over \lambda^2}\right)$ \\
$\lambda=1$ & ${22 \over 27}$ & $-{5 \over 81}$
& ${61 \over 6\cdot 81}$ \\
\hline
$\Xi$ & $-{2 \over 9}$
& $-{1 \over 81}\left(1+{14 \over \lambda}
+{6 \over \lambda^2}\right)$
& $-{1 \over 2\cdot 81}\left(1+{18 \over \lambda}
-{6 \over \lambda^2}\right)$ \\
$\lambda=1$ &{\bf $-{2\over 9}$} & $-{7 \over 27}$ & $-{13 \over 2\cdot 81}$ \\
\hline
$\left( \begin{array}{c}
\Lambda\hbox{-}\Sigma \\
\Sigma\hbox{-}\Lambda \\
\end{array} \right)$
& $\begin{array}{c}
-{2 \over 27}\left(7+{2 \over \lambda}\right) \\
-{2 \over 3} \\
\end{array}$
& $\begin{array}{c}
-{1 \over 81}\left(5-{2 \over \lambda}\right) \\
-{1 \over 27} \\
\end{array}$
& $\begin{array}{c}
-{1 \over 2\cdot 81}\left(13+{6 \over \lambda}\right) \\
-{19 \over 2\cdot 81} \\
\end{array}$ \\
\hline
\end{tabular}
\end{center}
\end{table}
%

\subsection{Born approximation}

Let us calculate the Scheerbaum factor
for the s.s. symmetric nuclei in the Born
approximation. For $B=N$ with $I_B=1/2$,
we have two possible isospin values $I=0$ and 1 in \eq{fm22}.
Then the invariant parts of the Born kernel in \eq{fm9} yield
\begin{eqnarray} 
S_N & = & {1 \over 8\pi} \left\{~\sum_\CT
(X_\CT^{LS})_{NN}^{I=0}
\left[\overline{f_\CT^{LS}(\theta)-f_\CT^{LS}(\pi-\theta)}\right]
\right. \nonumber \\
& & \left. +3 \sum_\CT (X_\CT^{LS})_{NN}^{I=1}
\left[\overline{f_\CT^{LS}(\theta)+f_\CT^{LS}(\pi-\theta)}\right]
\right\}\ \ .
\label{fm23}
\end{eqnarray} 
Here $I=0$ corresponds to the $\hbox{}^3E$ state
and $I=1$ to the $\hbox{}^3O$ state.
If we assume
\begin{equation}
\overline{f_\CT^{LS}(\theta)}=\overline{f_\CT^{LS}(\pi-\theta)}\ \ ,
\label{fm24}
\end{equation} 
we find that only the $\hbox{}^3O$ state contributes to $S_N$ and
obtain
\begin{equation}
S_N={3 \over 4\pi} \sum_\CT (X_\CT^{LS})_{NN}^{I=1}
~\overline{f_\CT^{LS}(\theta)}\ \ .
\label{fm25}
\end{equation} 
This expression corresponds to Scheerbaum's Eq.\,(3.57) \cite{SC76}.
Thus we have a correspondence
\begin{equation}
\sum_\CT (X_\CT^{LS})_{NN}^{I=1}~\overline{f_\CT^{LS}(\theta)}
\sim {4\pi \over {\bar q}} \int_0^\infty s^3~j_1({\bar q}s)
~g^{\hbox{}^3O}(s)~d\,s
\ \ .
\label{fm26}
\end{equation} 
Under the same assumption as \eq{fm24} the Scheerbaum factors
for the hyperons are given by
\begin{equation}
S_B={1 \over 2\pi}~{\xi \over 1+\xi}
\sum_\CT X_\CT^B~\overline{f_\CT^{LS}(\theta)}\ \ ,
\label{fm27}
\end{equation} 
where $X_\CT^B$ for $B=Y$ are defined by
\begin{eqnarray} 
& & X_\CT^B \nonumber \\
& & = \sum_I {2I+1 \over 2I_B+1} \left[
(X_\CT^{LS})_{BB}^{ud}+(X_\CT^{LS})_{BB}^{s}
+(X_\CT^{LS^{(-)}})_{BB}^{ud}
+(X_\CT^{LS^{(-)}\sigma})_{BB}^{s} \right].
\label{fm28}
\end{eqnarray} 
In this notation, $X_\CT^N$ is given by ($\xi$=1)
\begin{equation} 
X_\CT^N=3~(X_\CT^{LS})_{NN}^{I=1}=3~(X_\CT^{LS})_{NN}^{\hbox{}^3O}\ \ .
\label{fm29}
\end{equation} 
The spin-flavor factors $X_\CT^B$ can be easily obtained
from the explicit expressions
of $X_\CT^{LS}$, $X_\CT^{LS^{(-)}}$ and $X_\CT^{LS^{(-)}\sigma}$,
which are given in Appendix C of \cite{FU97}.\footnote{$\lambda
=m_{ud}/m_s$ in Appendix C of \protect\cite{FU97} is
a misprint of $\lambda=m_s/m_{ud}$.}
They are tabulated in Table 1.
When we derive these results, we should note
that $\CP^\prime=1$ in
\begin{equation} 
(X_\CT^{LS})_{a \CP,a^\prime \CP^\prime}=(X_\CT^{LS})_{a a^\prime}^{ud}
+(X_\CT^{LS})_{a a^\prime}^{s}~\CP^\prime
\label{fm30}
\end{equation} 
corresponds to the $\hbox{}^3O$ contribution.
For $B=\Sigma$, the isospin sum in \eq{fm28} gives
\begin{eqnarray} 
\hbox{isoscalar~term} & = & \sum_I {2I+1 \over 2I_B+1}\cdot 1
={1 \over 2I_B+1}~\sum_{II_z}~1 \nonumber \\
& = & {1 \over 2I_B+1}~\sum_{I_{Bz}}~1~\sum_\tau~1=2\ \ ,\nonumber \\
\hbox{isovector~term} & = & \sum_I {2I+1 \over 2I_B+1}
~(\bftau_1 \cdot \bftau_2)_I
={1 \over 2I_B+1}~\sum_{II_z} \langle\,II_z\,|\,\bftau_B \cdot
\bftau_N\,|\,II_z\,\rangle \nonumber \\
& = & {1 \over 2I_B+1}~(Tr\,\bftau_B)~(Tr\,\bftau_N) = 0\ \ .
\label{fm31}
\end{eqnarray} 
The factor 2 for the isoscalar term is the sum
over the proton and neutron, and the isovector
term does not contribute since we have
assumed $Z=N$ (the total isospin is zero
for $\alpha$, $\hbox{}^{16}\hbox{O}$ and $\hbox{}^{40}\hbox{Ca}$ ).

%

\bigskip
\begin{table}[t]
\caption{Quark-model parameters}
\label{table2}
\bigskip
%
\begin{center}
\renewcommand{\arraystretch}{1.4}
\setlength{\tabcolsep}{2mm}
\begin{tabular}{ccccc}
\hline
model & $b$ (fm) & $m_{ud}$ (MeV/$c^2$) & $\alpha_S$
& $\lambda=m_s/m_{ud}$ \\
\hline
RGM-F & 0.6   & 313 & 1.5187 & 1.25  \\
FSS   & 0.616 & 360 & 2.1742 & 1.526 \\
RGM-H & 0.667 & 389 & 2.1680 & 1.490 \\
\hline
\end{tabular}
\end{center}
\end{table}
%

Next we discuss the averaging procedure
in $\overline{f_\CT^{LS}(\theta)}$.
A possible approximation to
obtain $\overline{f_\CT^{LS}(\theta)}$ is
to use the Wigner transform $G_W^{LS}(\br, \bp)$ (which is
given in Eq.\,(2.16) of \cite{FU97}) at $\bp=0$,
and to follow the Scheerbaum's
prescription for the local potential $G_W^{LS}(\br, 0)$. We can
show that this procedure is equivalent
to set $\bq=0$ in \eq{a3}.
In this case the $\theta$-dependence in $f_\CT^{LS}(\theta)$
disappears ($\theta=\pi$) and $\overline{f_\CT^{LS}(\theta)}$ becomes
a function of the momentum transfer $k=|\bk|$ ($\bk=\bq_f-\bq_i$).
This corresponds to the Scheerbaum's parameter $\overline{q}$.
We call this the Scheerbaum approximation.

The relationship between the basic Born kernel in \eq{fm5} and
the Wigner transform for the $LS$ component is given by
\begin{eqnarray}
M(\bq_f, \bq_i) & = & \int d \br~e^{-i \bk \cdot \br}~G_W(\br, \bq)
\nonumber \\
& = & \sum_{\CT} X_{\CT}^{LS} \int d \br~e^{-i \bk \cdot \br}
~G_{W \CT}^{LS}(\br, \bq)~[\,\br \times \bq\,] \cdot \bS\ \ ,
\label{fm32}
\end{eqnarray} 
where $\bk=\bq_f-\bq_i$ and $\bq=(1/2)(\bq_f+\bq_i)$.
The spatial function $G_{W \CT}^{LS}(\br, \bq=0)$ becomes
a function of $r=|\br|$ only, and thus we can carry out the
angle integral $\int d \hat{\br}$ after the partial wave
expansion of the plane wave. Then the component with the angular
momentum $\ell=1$ only survives, and we obtain
\begin{eqnarray}
& & M(\bq_f, \bq_i) |_{\bq=0} \nonumber \\
& & = \sum_{\CT} X_{\CT}^{LS}~{4\pi \over k}
\int_0^\infty r^3~ d\,r~j_1(kr)~G_{W\CT}^{LS}(\br, 0)
~\CO^{LS}(\bq_f, \bq_i),
\label{fm33}
\end{eqnarray} 
or
\begin{equation} 
f_\CT^{LS}(\theta) |_{\bq=0}={4\pi \over k}
\int_0^\infty r^3~ d\,r~j_1(kr)~G_{W\CT}^{LS}(\br, 0)\ \ .
\label{fm34}
\end{equation} 
Note that $|_{\bq=0}$ implies setting $\bq=0$ except for
the $LS$ operator part. If we call $G_{W}^{LS}(\br, 0)
=\sum_\CT X_{\CT}^{LS} G_{W\CT}^{LS}(\br, 0)$ the $LS$ potential,
we find 
\begin{equation} 
\sum _\CT X_{\CT}^{LS} f_\CT^{LS}(\theta) |_{\bq=0}={4\pi \over k}
\int_0^\infty r^3~ d\,r~j_1(kr)~G_{W}^{LS}(\br, 0)\ \ ,
\label{fm35}
\end{equation} 
which is nothing but \eq{fm26} if we assign
\begin{equation} 
G_{W}^{LS}(\br, 0) \sim g^{\hbox{}^3O}(r) \ \ .
\label{fm36}
\end{equation} 
Then we find 
\begin{equation}
\overline{f_\CT^{LS}(\theta)} \sim f_\CT^{LS}(\theta) |_{\bq=0}\ \ ,
\label{fm37}
\end{equation} 
with $k=\overline{q}$ (see \eq{a5}).
We will discuss the choice of the value $k=\overline{q}$ and
a further simplification in the next section,

Another approximation for the $LS$ function $f_\CT^{LS}(\theta)$ in
\eq{fm27} is obtained by taking only $P$-wave components
in the partial wave expansion of the Born kernel \cite{LS99}.
Suppose the partial wave expansion of \eq{fm8} is
\begin{eqnarray}
& & \sum_\Omega
M_{aa^\prime}^{\Omega~\hbox{total}} (\bq_f, \bq_i)
~\CO^\Omega(\bq_f, \bq_i) \nonumber \\
& & =\sqrt{(1+\delta_{a_1, a_2})(1+\delta_{a_1^\prime, a_2^\prime})}
\sum_{JM\ell \ell^\prime SS^\prime} 4\pi
~R_{\alpha S\ell, \alpha^\prime S^\prime \ell^\prime}
^{\Omega~J}(q_f, q_i)\nonumber \\
& & \times \CY_{(\ell S)JM}(\hat{\bq}_f; spin)
~\CY_{(\ell^\prime S^\prime)JM}^*(\hat{\bq}_i; spin)\ \ ,
\label{fm38}
\end{eqnarray} 
where $~\CY_{(\ell S)JM}(\hat{\bq}; spin)
=[Y_\ell(\hat{\bq})\chi_S(spin)]_{JM}$ is
the standard space-spin wave function.
The front factor $\sqrt{(1+\delta_{a_1, a_2})
(1+\delta_{a_1^\prime, a_2^\prime})}$ is 2
for $NN$ and 1 for $YN$.
The partial-wave amplitudes $R_{\alpha S\ell,
\alpha^\prime S^\prime \ell^\prime}^{\Omega~J}
(q_f, q_i)$ for $\Omega=LS$ and $LS^{(-)}$ are explicitly given by
\begin{eqnarray}
& & R_{\alpha S\ell, \alpha^\prime S^\prime \ell^\prime}
^{LS~J}(q_f, q_i)=\delta_{\ell, \ell^\prime}~\delta_{S, S^\prime}
~\delta_{S, 1}
~q_f q_i~{1 \over 2(2\ell+1)}~[\ell (\ell+1)+2-J(J+1)]\nonumber \\
& & \times \sum_\CT \left(X_\CT^{LS}\right)_{\alpha \alpha^\prime}
\left(f_{\CT~\ell+1}^{LS}-f_{\CT~\ell-1}^{LS}\right)\ \ ,\nonumber \\
& & R_{\alpha S\ell, \alpha^\prime S^\prime \ell^\prime}
^{LS^{(-)}~J}(q_f, q_i)=\delta_{\ell, \ell^\prime}~\delta_{J, \ell}
~q_f q_i {\sqrt{J(J+1)} \over 2J+1}\sum_\CT
\left[~\left(X_\CT^{LS^{(-)}}\right)
_{\alpha \alpha^\prime}\right. \nonumber \\
& & \left. +\left(X_\CT^{LS^{(-)}\sigma}\right)
_{\alpha \alpha^\prime}~(-1)^{1-S^\prime}~\right]
~\left(f_{\CT~\ell-1}^{LS}-f_{\CT~\ell+1}^{LS}\right)\nonumber \\
& & \ \ \hspace{70mm} (S, S^\prime=1, 0~\hbox{or}
~0, 1~\hbox{only})\ ,
\label{fm39}
\end{eqnarray} 
where
\begin{equation}
f_{\CT \ell}^{LS}={1 \over2} \int_0^\pi f_\CT^{LS}(\theta)
~P_\ell(\cos \theta)
~\sin \theta d~\theta
\label{fm40}
\end{equation} 
is the angular-momentum projection of the $LS$ function \eq{a3}.
(See Eq.\,(2.29) of \cite{LS99}.)
If we take $\ell=\ell^\prime=1$ only
in \eq{fm38} and use the formulae ($\bn=[\bq_i \times
\bq_f]$)
\begin{eqnarray}
\CO^{LS} & = & i \bn \cdot \bS=-{2\pi \over 3}q_f q_i
\sum_{JM} [4-J(J+1)]\nonumber \\
& & \times \CY_{(11)JM}(\hat{\bq}_f; spin)
~\CY_{(11)JM}^*(\hat{\bq}_i; spin)\ \ ,\nonumber \\
\CO^{LS^{(-)}} & = & i \bn \cdot \bS^{(-)}
=4\pi {\sqrt{2} \over 3}q_f q_i \sum_{M}
\left\{~\CY_{(11)1M}(\hat{\bq}_f; spin)
~\CY_{(10)1M}^*(\hat{\bq}_i; spin)\right. \nonumber \\
& & \left. +\CY_{(10)1M}(\hat{\bq}_f; spin)
~\CY_{(11)1M}^*(\hat{\bq}_i; spin)\right\}
\ \ ,\nonumber \\
\CO^{LS^{(-)}\sigma} & = & i \bn \cdot \bS^{(-)}
P_\sigma=4\pi {\sqrt{2} \over 3} q_f q_i
\sum_{M} \left\{~-\CY_{(11)1M}(\hat{\bq}_f; spin)
~\CY_{(10)1M}^*(\hat{\bq}_i; spin)\right. \nonumber \\
& & \left. +\CY_{(10)1M}(\hat{\bq}_f; spin)
~\CY_{(11)1M}^*(\hat{\bq}_i; spin)\right\}
\ \ ,
\label{fm41}
\end{eqnarray} 
we eventually find that this prescription corresponds to the
approximation\footnote{Note that $\CO^{LS^{(-)}\sigma}$ part
disappears because of the second equation of \eq{fm14}.}
\begin{eqnarray}
M_{aa^\prime}^{LS~\hbox{total}} (\bq_f, \bq_i)
& \sim & \left\{
\begin{array}{c}
2 \sum_\CT \left(X_\CT^{LS}\right)_{NN}^{I=1}
~\left(f_{\CT 0}^{LS}-f_{\CT 2}^{LS}\right)  \\
\sum_\CT \left[~\left(X_\CT^{LS}\right)_{aa^\prime}^{ud}
+\left(X_\CT^{LS}\right)_{aa^\prime}^{s}~\right]
~\left(f_{\CT 0}^{LS}-f_{\CT 2}^{LS}\right)  \\
\end{array}
\right. \nonumber \\
& & \ \ \hspace{40mm} \hbox{for} \quad \left\{ \begin{array}{c}
NN \\
YN \\
\end{array}
\right. \ \ ,\nonumber \\
M_{aa^\prime}^{LS^{(-)}~\hbox{total}} (\bq_f, \bq_i)
& \sim & \sum_\CT \left[~\left(X_\CT^{LS^{(-)}}\right)_{aa^\prime}^{ud}
+\left(X_\CT^{LS^{(-)}\sigma}\right)_{aa^\prime}^{s}~\right]
\nonumber \\
& & \times \left(f_{\CT 0}^{LS}-f_{\CT 2}^{LS}\right)
\qquad \hbox{for} \quad YN \ \ .
\label{fm42}
\end{eqnarray} 
If we compare these with \eq{fm9}, we find that this approximation
corresponds to setting
\begin{equation}
\overline{f_\CT^{LS}(\theta)} \sim f_{\CT 0}^{LS}
-f_{\CT 2}^{LS}\ \ .
\label{fm43}
\end{equation} 
Note that we still have two parameters $q_f=|\bq_f|$ and $q_i=|\bq_i|$,
the choice of which will be discussed in the next section.

%
\bigskip
\begin{table}[b]
\caption{Contributions of symmetric ($sLS$) and antisymmetric 
($aLS$) $LS$ terms of the FB interaction \protect\cite{FU97} to
the nucleon Scheerbaum factor $S_N$ in the simplest approximation
with ${\bar q}=0$. The unit is $\hbox{MeV}\cdot \hbox{fm}^5$.}
\label{table3}
\bigskip
\begin{center}
\renewcommand{\arraystretch}{1.4}
\setlength{\tabcolsep}{2mm}
\begin{tabular}{ccccc}
\hline
model & $sLS$  & $aLS$  & $S_N$ & $aLS/sLS$ \\
 & $D_-+D_+$ & $S+S^\prime$ & total & ratio      \\
\hline
RGM-F & $-28.2$   & $-9.7$  & $-37.8$ & 0.344  \\
FSS   & $-32.2$   & $-11.0$ & $-43.2$ & 0.342  \\
RGM-H & $-32.2$   & $-11.0$ & $-43.2$ & 0.342  \\
\hline
\end{tabular}
\end{center}
\end{table}
%

\subsection{Realistic calculation}

Here we consider a realistic calculation of $S_B$,
based on the $G$-matrices
for the $NN$ and $YN$ interaction \cite{KO99}.
In this case, $M_{BB}^{\Omega~\hbox{total}}(\bk_{12},
\bk_{12}^\prime)$ in \eq{fm16} now depends
on some averaged value $K$ of $\sqrt{(\bK_{12})^2}$ and
the starting energy $\omega=E_B(k_1^\prime)+E_N(k_2^\prime)$; 
i.e., $G_{BB}^{\Omega~\hbox{total}}(\bk_{12}, \bk_{12}^\prime;
K, \omega)$. We assume the density $\rho_\tau
(\bk_2, \bk_2^\prime)$ in \eq{fm15} is well approximated by
the local form
\begin{equation}
\rho_\tau(\bk_2, \bk_2^\prime)=\delta(\bk_2-\bk_2^\prime)
~\rho_\tau(\bk_2)\ \ ,
\label{fm44}
\end{equation}
where $\rho_\tau(\bk)$ is given by the Fermi distribution
\begin{equation}
\rho_\tau(\bk)=2~\Theta(k_F^\tau-|\bk|)\ \ ,
\label{fm45}
\end{equation}
with $\Theta$ being the Heaviside's step function.
Under this assumption, we can approximately
set $\bk_2 \sim \bk_2^\prime=\bk_1+\bk_2-\bk_1^\prime$.
This implies  $\bk_1=\bk_1^\prime$ and $\bk_2=\bk_2^\prime$.
This approximation, however, makes
the $\ell s$ factor in \eq{fm18} disappear.
Thus we must take the density average
by Eqs.\,{(\ref{fm44}) and (\ref{fm45}) only
for the invariant part just as in Eq.\,{(\ref{fm21}) and (\ref{fm22}).
After the change of the notation $\bk_2 \rightarrow \bq_2$ and
$\bk_{12} \rightarrow \bq$ {\em etc.}, the expression we use is 
\begin{eqnarray}
& & \overline{G_{BB}^{\Omega}(\bk_{12}, \bk_{12}^\prime;
K, \omega)} =
{\int \rho_\tau(\bq_2)~G_{BB}^{\Omega}(\bq, \bq^\prime;
K, \omega)~d\,\bq_2 \over \int \rho_\tau(\bq_2)~d\,\bq_2}
\nonumber \\
& & = {3 \over 4\pi}~{1 \over (k_F)^3}
\int_{|\bq_2|<k_F}
G_{BB}^{\Omega}(\bq, \bq^\prime;K, \omega)~d\,\bq_2\ \ ,
\label{fm46}
\end{eqnarray} 
where we have assumed symmetric nuclear matter and
used $\rho_\tau(\bq_2) \rightarrow \Theta(k_F-|\bq_2|)$.
Here we make use of the same treatment of angle-averaging just as used
for the derivation of the s.p. potentials in \cite{KO99}.
We first change the integral variable $\bq_2$ in \eq{fm46}
to $\bq$ through $\bq_2=\xi \bq_1-(1+\xi) \bq$.
We assume $q_1$ and determine $K$ through
$K=\sqrt{\overline{(\bK_{12})^2}(q_1,q)}$.
Since $q_2=|\bq_2|$ is given by $q_1$ and $K$,
it is also determined from $q_1$ and $q$.
The starting energy $\omega$ is
determined as $\omega=E_B(q_1)+E_N(q_2)$. Then by using
the weight function $W(q_1, q)$ introduced in Eq.\,(21) of
\cite{KO99}, we find
\begin{eqnarray}
\overline{G_{BB}^{\Omega}(\bk_{12}, \bk_{12}^\prime;
K, \omega)}& = & {3 \over 4\pi}~{1 \over (k_F)^3}~(1+\xi)^3
\int_0^{q_{max}} q^2~d\,q~W(q_1, q) \nonumber \\
& & \times \int d\,\widehat{\bq}
~G_{BB}^{\Omega}(\bq, \bq; K, \omega)\ \ .
\label{fm47}
\end{eqnarray} 
%

%
\begin{table}[t]
\caption{The Scheerbaum factor $S_B$ in the
simplest ${\bar q}=0$ approximation
by the $\protect\bp=0$ Wigner transform $G_W(0)$.
In $S_{\Lambda\hbox{-}\Sigma}=S_{\Sigma\hbox{-}\Lambda}$,
the average mass of $\Lambda$ and $\Sigma$ is used for $\xi$.
The unit is $\hbox{MeV}\cdot \hbox{fm}^5$.
$\lambda=(m_s/m_{ud})$ implies the FSB.
When $\lambda=1$, we also assume $\xi=1$.}
\label{table4}
\bigskip
%
\begin{center}
\renewcommand{\arraystretch}{1.4}
\setlength{\tabcolsep}{2mm}
\begin{tabular}{ccrrcrrcrr}
\hline
$B$ & & \multicolumn{2}{c}{RGM-F} & &\multicolumn{2}{c}{FSS}
 &  & \multicolumn{2}{c}{RGM-H} \\
\cline{3-4} \cline{6-7} \cline{9-10}
 &  & $\lambda=1$ & $\lambda=1.25$ & & $\lambda=1$
    & $\lambda=1.526$ & & $\lambda=1$ & $\lambda=1.490$ \\
\hline
  $N$     & & $-37.8$ &        & & $-43.2$ &         & & $-43.2$ &   \\
$\Lambda$ & & $-12.4$ & $-9.0$ & & $-14.1$ &  $-8.3$ & & $-14.1$
& $-8.6$ \\
$\Sigma$  & &$-22.7$ & $-19.3$ & & $-26.0$ & $-21.5$ & & $-26.0$
& $-21.5$ \\
$\Xi$     & & 12.3   &  9.2    & & 14.0    &  9.5    & & 14.0  
&  9.7    \\
$S_{\Lambda\hbox{-}\Sigma}$ & & 20.5    & 16.8  & & 23.4    & 18.5
& & 23.5    &  18.6   \\ 
\hline
\end{tabular}
\end{center}
\end{table}
%

In order to reduce the angular integral in \eq{fm47} further,
we need explicit formulae for the partial-wave decomposition
of the invariant amplitudes.\footnote{The full
expression of the partial-wave decomposition
of the invariant amplitudes is given in Appendix D
of \cite{LS99}.}
For the $LS$ and $LS^{(-)}$ components,
these are given by $M_{aa^\prime}^{LS~\hbox{total}}(\bq_f, \bq_i)
=(2/|\bn|)~h_{aa^\prime}^0$ and $M_{aa^\prime}^{LS^{(-)}~\hbox{total}}
(\bq_f, \bq_i)=(2/|\bn|)~h_{aa^\prime}^-$, where $h_{aa^\prime}^0$
and $h_{aa^\prime}^-$ are the flavor matrix elements of
\begin{eqnarray}
& & h^0 = -{1 \over 4} \sum_J {(2J+1) \over J(J+1)}
\left[~G_{1J, 1J}^J~P_J^1(\cos \theta)\right. \nonumber \\
& & \left. -(J+1)~G_{1\,J-1,\,1\,J-1}^J~P_{J-1}^1(\cos \theta)
+J~G_{1\,J+1,\,1\,J+1}^J~P_{J+1}^1(\cos \theta)
~\right]\ \ ,\nonumber \\
& & h^- = {1 \over 4} \sum_J {2J+1 \over \sqrt{J(J+1)}}
~\left[~G_{1J, 0J}^J+G_{0J, 1J}^J~\right]~P_J^1(\cos \theta)
\ \ .
\label{fm48}
\end{eqnarray} 
Here $G_{S \ell, S^\prime \ell^\prime}^J=G_{S \ell, S^\prime
\ell^\prime}^J (q_f, q_i; K, \omega)$ and $P_J^1(\cos \theta)$ is
the associated Legendre function of the first kind
with degree 1. 
We set $\bq_f=\bq_i=\bq$ and $\theta=0$, and use $(1/\sin \theta)
P_J^1(\cos \theta)={P_J}^\prime(\cos \theta)$ and ${P_\ell}^\prime(1)
=\ell(\ell+1)/2$. Then we easily find
\begin{eqnarray}
& & G^{LS}(\bq, \bq; K, \omega) = -{1 \over 4q^2}
\sum_{\ell=1}^\infty \left[~(2\ell-1)(\ell+1)
~G_{1\ell, 1\ell}^{\ell-1}\right. \nonumber \\
& & \left. +(2\ell+1)~G_{1\ell, 1\ell}^\ell
-(2\ell+3)\ell~G_{1\ell, 1\ell}^{\ell+1}~\right]\ \ ,\nonumber \\
& & G^{LS^{(-)}}(\bq, \bq; K, \omega) = {1 \over 4q^2}
\sum_{\ell=1}^\infty (2\ell+1)\sqrt{\ell (\ell+1)}
~\left[~G_{1\ell, 0\ell}^\ell+G_{0\ell, 1\ell}^\ell~\right].
\label{fm49}
\end{eqnarray} 
Combining Eqs.\,(\ref{fm22}), (\ref{fm47}) and (\ref{fm49}),
we obtain
\begin{eqnarray}
& & S_B(q_1) = -(1+\delta_{B,N})~{1 \over 2\pi}
~{3 \over 4(k_F)^3}~\xi(1+\xi)^2 
\sum_{I,J} {2I+1 \over 2I_B +1} (2J+1)\nonumber \\
& & \times \int_0^{q_{max}} d\,q~W(q_1, q)
~\left\{~(J+2)G^J_{B1\,J+1,\,B1\,J+1}(q, q; K, \omega)
\right. \nonumber \\ 
& & +G^J_{B1J, B1J}(q, q; K, \omega)
-(J-1) G^J_{B1\,J-1,\,B1\,J-1}(q, q; K, \omega) \nonumber \\
& & \left. -\sqrt{J(J+1)}~\left[~G^J_{B1J, B0J}(q, q; K, \omega)
+G^J_{B0J, B1J}(q, q; K, \omega) \right] \right\}\ \ .
\label{fm50}
\end{eqnarray} 
%


\begin{table}[t]
\caption{$LS$ and $LS^{(-)}+LS^{(-)}\sigma$ contributions to the 
Scheerbaum factor $S_\Lambda$
in the simplest approximation of \protect\eq{re2}.
The model is FSS. $\lambda=(m_s/m_{ud})$ implies the FSB.
When $\lambda=1$, we also assume $\xi=1$.}
\label{table5}
\bigskip
\begin{center}
\renewcommand{\arraystretch}{1.4}
\setlength{\tabcolsep}{2mm}
\begin{tabular}{crrrrr}
\hline
& $X_{D_-}^\Lambda$ & $X_{D_+}^\Lambda$
& $X_{S}^\Lambda$ & $\widetilde{S}_\Lambda$
& $S_\Lambda$ \\
\hline
($\lambda=1$) & & & & & \\
$LS$ & ${8 \over 9}$ & $-{2 \over 9}$ & ${1 \over 9}$
& 0.998 & $-24.4$ \\ 
$LS^{(-)}+LS^{(-)}\sigma$ & $-{2 \over9}$ & $-{1 \over 9}$
& $-{1 \over 18}$ & $-0.421$ & 10.3 \\
\hline
sum & ${2 \over 3}$ & $-{1 \over 3}$ & ${1 \over 18}$
& 0.577 & $-14.1$ \\
\hline
($\lambda=1.526$) & & & & & \\
$LS$ & 0.6723 & $-0.1395$ & 0.1014 & 0.813 & $-18.2$ \\
$LS^{(-)}+LS^{(-)}\sigma$ & $-0.2856$ & $-0.0539$
& $-0.0525$ & $-0.440$ & 9.9 \\
\hline
sum & 0.3867 & $-0.1934$ & 0.0489 & 0.373 & $-8.3$ \\
\hline
\end{tabular}
\end{center}
\end{table}
%

\section{Result and discussion}

For the value of $k=\overline{q}$ in \eq{fm37},
we follow the suggestion by
Scheerbaum \cite{SC76} and take the value corresponding
to the "wavelength" of the density distribution.
If we take this to be $\sim 4t$ with $t\sim 2.4~\hbox{fm}$ being
the nuclear surface thickness, we arrive
at the estimate\footnote{This is almost half of the Fermi
momentum $k_F=(9\pi/8)^{1/3}/r_0
=1.36~\hbox{fm}^{-1}$ for $r_0=1.12~\hbox{fm}$,
which corresponds to $\rho=(3/4\pi)/{r_0}^3=0.170~\hbox{fm}^{-3}$.}
\begin{equation}
k=\overline{q} \sim {2\pi \over 4t} \sim 0.7~\hbox{fm}^{-1}
\ \ .
\label{re1}
\end{equation}
The simplest approximation is
to set $k=\overline{q}=0$ in \eq{a5}.
In this case we can write down an analytic expression for $S_B$ :
\begin{eqnarray}
S_B & = & -\alpha_S\,x^3\,m_{ud}c^2\,b^5
~{\xi \over 1+\xi}~\widetilde{S}_B\ \ ,\nonumber \\
\widetilde{S}_N & = & {14 \over 9}
-{10 \over 27} \left({3 \over 4}\right)^{{3 \over 2}}
+{32 \over 81} \left({12 \over 11}\right)^{{3 \over 2}}\ \ ,\nonumber \\
\widetilde{S}_\Lambda & = & {2 \over 9\lambda}
\left( 2+{1\over \lambda}\right)
-{1 \over 9\lambda} \left( 2+{1\over \lambda}\right)
\left({3 \over 4}\right)^{{3 \over 2}}
+{1 \over 9\lambda} \left(2-{1\over \lambda}\right)
\left({12 \over 11}\right)^{{3 \over
2}}\ \ ,\nonumber \\
\widetilde{S}_\Sigma & = & {2 \over 3 \cdot 81}\left(
106-{6 \over \lambda}-{1 \over \lambda^2} \right)
- {1 \over 81} \left(18-{10 \over \lambda}-{3 \over \lambda^2} \right)
\left({3 \over 4}\right)^{{3 \over 2}} \nonumber \\
& & +{1 \over 3 \cdot 81}\left(26+{24 \over \lambda}
-{7 \over \lambda^2} \right)
\left({12 \over 11}\right)^{{3 \over 2}}\ \ ,\nonumber \\
\widetilde{S}_\Xi & = & -{2 \over 9}-{1 \over 81} \left(1+
{14 \over \lambda}+{6 \over \lambda^2} \right)
\left({3 \over 4}\right)^{{3 \over 2}}
-{1 \over 81} \left(1+{18 \over \lambda}-{6 \over \lambda^2} \right)
\left({12 \over 11}\right)^{{3 \over 2}}\ \ ,\nonumber \\
\widetilde{S}_{\Lambda \hbox{-} \Sigma}
& = & -{2 \over 27} \left(7+{2 \over \lambda}\right)
-{1 \over 81} \left(5-{2 \over \lambda} \right)
\left({3 \over 4}\right)^{{3 \over 2}}
-{1 \over 81} \left(13+{6 \over \lambda} \right)
\left({12 \over 11}\right)^{{3 \over 2}},
\label{re2}
\end{eqnarray}
where $x=(\hbar/m_{ud}cb)$, $\lambda=(m_s/m_{ud})$ and $\xi=(M_N/M_B)$.
The value of $S_B$ in this simplest approximation is given in Table 4.
The parameters of our three models, RGM-F, FSS and RGM-H,
are given in Table 2.
We note that FSS and RGM-H produce very similar results for 
the s.p. $\ell s$ force, because the strength
factor, $\alpha_S\,x^3\,m_{ud}c^2\,b^5\,\xi/(1+\xi)$,  
and the quark-mass ratio, $\lambda=(m_s/m_{ud})$, are very
similar to each other. In particular,
$\sqrt{2/\pi}\,\alpha_S\,x^3\,m_{ud}c^2$ is constrained
to be $440~\hbox{MeV}$ for RGM-F and FSS,
in order to reproduce the $N$-$\Delta$ mass splitting through
the color-magnetic term of the FB interaction.
The $S_N$ ($=S_N^{\hbox{}^3O}$) value, $\sim -40~\hbox{MeV}\cdot
\hbox{fm}^5$, is rather small,
compared with the $S^{\hbox{}^3O}_{free}$ value,
$-53 \sim -61~\hbox{MeV}\cdot \hbox{fm}^5$,
given in Table 1 of the Scheerbaum's paper \cite{SC76} for
the Reid soft-core and other potentials.
In his calculation, the effect of the short-range correlation
reduces this value
to $-34 \sim -47~\hbox{MeV}\cdot \hbox{fm}^5$ in the same Table 1.
These values were obtained with $\overline{q}=0.7~\hbox{fm}^{-1}$ in
the Scheerbaum approximation.
However, we will see in the following
that the effect of the short-range
correlation obtained by solving the $G$-matrix equation
is very small in our case.
This is probably because our short-range repulsion
is not represented by the hard core
but by the quark-exchange kernel.

%
\begin{table}[t]
\caption{The Scheerbaum factors $S_B$ predicted
by FSS in various types of approximations.
1) $G_W(0)$: Born approximation with $\protect\bp=0$ Wigner
transform with ${\bar q}=0$,
2) $G_W$: Born approximation with $\protect\bp=0$ Wigner
transform with ${\bar q}=0.7~\hbox{fm}^{-1}$,
3) $P$: $P$-wave Born approximation with $q_f=q_i=0.35~\hbox{fm}^{-1}$,
4) $G$-matrix: $QTQ$ or continuous choice
with $q_1=0$ and $k_F=1.35~\hbox{fm}^{-1}$.
The unit is $\hbox{MeV}\cdot \hbox{fm}^5$.
}
\label{table6}
\bigskip
\begin{center}
\renewcommand{\arraystretch}{1.4}
\setlength{\tabcolsep}{4mm}
\begin{tabular}{ccrrrcrrr}
\hline
$B$ & & \multicolumn{3}{c}{Born} & & \multicolumn{3}{c}{$G$-matrix} \\
\cline{3-5} \cline{7-9}
    & & $G_W(0)$ & $G_W$ & $P$ \quad & & $QTQ$ & cont. & ratio \\
\hline
    $N$   & & $-43.2$ & $-40.5$ & $-41.7$ & & $-40.4$ & $-41.6$ & 1 \\
$\Lambda$ & & $-8.3$  & $-7.8$  & $-8.0$  & & $-3.8$ & $-3.4$
          & $\sim {1 \over 12}$ \\
$\Sigma$  & & $-21.5$ & $-20.1$ & $-20.7$ & & $-27.5$ & $-22.4$
          & $\sim {1 \over 2}$ \\
$\Xi$     & & 9.5 &  9.0 &  9.2 & &  &  &  \\

\hline
\end{tabular}
\end{center}
\end{table}
%

Table 3 shows that the Galilean
non-invariant $aLS$ term of the FB
interaction \cite{FU97} has a fairly large contribution 
to $S_N^{\hbox{}^3O}$. The magnitude of $aLS$
contribution is almost 1/3 of the $sLS$ contribution
and they reinforce each other with
the same sign \cite{SU84}.

We note that the $S_\Lambda$ value changes significantly by
the FSB, which is easily understood
from the spin-flavor factors in Table 1.
All the $X_\CT^\Lambda$ factors contain the factor $1/\lambda$.
If we assume $S_N^{\hbox{}^3O}=1$,
then $S_\Lambda$ is about 1/3 for $\lambda=1$,
while it is $\sim 1/5$ for $\lambda
=1.69$ (the maximum FSB).\footnote{When we
set $\lambda=1$ in \protect\eq{re2}, we also neglect the mass
difference of baryons; i.e., $\xi=1$.} 
On the other hand, $S_\Sigma$ does not change very much
by the FSB: it changes from 3/5 to 1/2 as $\lambda$ changes
from 1 to 1.69. 
The sign of $S_\Xi$ is positive
and its value changes from $-1/3$ to $-1/4$.
The $\Lambda$-$\Sigma$ coupling term is not small,
and is about half of $S_N$ both
in $\lambda=1$ and $\lambda\neq 1$ cases.
The sign of $S_{\Lambda \hbox{-} \Sigma}
=S_{\Sigma \hbox{-} \Lambda}$ depends on
the phase convention of the $\Lambda$ and $\Sigma$ flavor functions.

Table 5 shows the decomposition
of $S_\Lambda$ into $LS$ and $LS^{(-)}+LS^{(-)}\sigma$
contributions in the simplest approximation. 
The signs of these two contributions
are opposite to each other,
and they largely cancel;
namely, the half of the $LS$ contribution is cancelled by
the $LS^{(-)}+LS^{(-)}\sigma$ contribution.
Because of this cancellation, the strong $\lambda$-dependence
in $S_\Lambda$ is even enhanced.

Table 6 shows the predictions of $S_B$ by FSS,
calculated in the various prescriptions.
The first column with $G_W(0)$ implies
the simplest $\overline{q}=0$ prescription,
the second with $G_W$ the Scheerbaum approximation
with $\overline{q}=0.7~\hbox{fm}^{-1}$,
the third with the $P$-wave approximation of \eq{fm43}.
In the last case, we have
assumed $q_f=q_i=Q$ with $Q=\overline{q}/2=0.35~\hbox{fm}^{-1}$. 
We have examined the $\overline{q}$ or $Q$ dependence
in \eq{fm37} or \eq{fm43}.
Actually, the averaged spatial
function $\overline{f_\CT^{LS}(\theta)}$ has some momentum
dependence, so does $S_B$. However, this weak momentum dependence
almost disappears if we take the ratio $S_B/S_N$.
After all, we have found that $S_B/S_N$ ratios
in the Born approximation are approximately given by
\begin{equation}
{S_\Lambda \over S_N} \sim {1 \over 5}\ \ ,\quad
{S_\Sigma \over S_N} \sim {1 \over 2}\ \ ,\quad
{S_\Xi \over S_N} \sim -{1 \over 4}\ \ ,
\label{re3}
\end{equation}
independently of whichever approximation
of the spatial functions is used.
Table 6 also shows the results of the realistic calculation
using $G$-matrix solutions in the $QTQ$ and continuous prescriptions
for intermediate spectra.
Here the $q_1$ value in $S_B(q_1)$ (see \eq{fm50}) is assumed
to be $q_1=0$.
We find that $S_\Lambda$ receives a strong effect due to the
short-range correlation and $S_\Lambda/S_N$ is
further reduced to almost 1/12.
On the other hand, $S_N$ and $S_\Sigma$ do not change so much,
except for the increase of $|S_\Sigma|$ in the $QTQ$ prescription.
The ratio, $S_\Sigma/S_N \sim 1/2$, does not seem to
change very much even in the $G$-matrix calculation.

%
\begin{table}[t]
\caption{The nuclear-matter density dependence of the 
Scheerbaum factors $S_B$ for $N$,
$\Lambda$ and $\Sigma$, predicted by quark-model $G$-matrices
in the continuous prescription.
The model is FSS. The unit is $\hbox{MeV}\cdot \hbox{fm}^5$.
}
\label{table7}
\bigskip
\renewcommand{\arraystretch}{1.4}
\setlength{\tabcolsep}{4mm}
\begin{center}
\begin{tabular}{@{}cc@{\extracolsep{\fill}}cccc}
\hline
$k_F$ (fm$^{-1}$) & & $1.07$ & $1.20$ & $1.35$ & \\
\hline
      $N$ &  & $-43.0$ & $-42.3$ & $-41.3$ & ( 1 ) \\
$\Lambda$ &  & $-2.0$  & $-2.7$  & $-3.5$  & ($\sim {1 \over 12}$) \\
$\Sigma$  &  & $-21.5$ & $-22.0$ & $-21.8$ & ($\sim {1 \over 2}$) \\
\hline
\end{tabular}
\end{center}
\end{table}
%


Table 7 shows the $k_F$ dependence of $S_B (q_1=0)$ for $N$,
$\Lambda$ and $\Sigma$,
which are calculated from the FSS $G$-matrices
with the continuous choice.\footnote{In \protect\cite{KO99}
we assumed $U_B (q_1)=U_B (q_1=3.8~\hbox{fm}^{-1})$ for $q_1
\geq 3.8~\hbox{fm}^{-1}$, in order to avoid the unrealistic
behavior \protect\cite{LS99} of the s.p. potentials
in the high momentum region.
The results in Tables 7 and 8 are obtained with this prescription,
while those in Table 6 are without this prescription.
The difference of $S_B$ between these two prescriptions
is very small, as is seen for $k_F=1.35~\hbox{fm}^{-1}$.} 
The three values of the Fermi momentum,
$k_F =1.07,\;1.2$ and 1.35 fm$^{-1}$, correspond to
the three densities of $\rho= 0.5\rho_0,\;0.7\rho_0$ and $\rho_0$,
respectively. Here $\rho_0=0.17~\hbox{fm}^{-3}$ is
the normal density.
We find that $S_\Lambda/S_N$ becomes even smaller
for lower densities, while $S_\Sigma/S_N$ does not change much.
Each contribution from the $LS$ and the $LS^{(-)}$ components
in even and odd states
as well as $I=1/2$ and $I=3/2$ channels is shown in Table 8
for $k_F=1.35~\hbox{fm}^{-1}$.
It is clear that in the case
of $S_{\Lambda}$ the $LS^{(-)}$ contribution
almost cancels the $LS$ one just as in the Born approximation,
which makes the ratio $S_{\Lambda} /S_N$ to be less than 1/10
for $\rho \lesssim \rho_0$.
For the $\Sigma$ hyperon, the contribution from the $LS^{(-)}$
force has an opposite sign to that for the $\Lambda$ hyperon,
and the ratio $S_{\Sigma} /S_N$ turns out to be about 1/2
even in the realistic calculation, which is very much independent
of the precise value of $k_F$.

%
\begin{table}[t]
\caption{Decomposition
of $S_{\Lambda} = -3.5$ MeV$\cdot$fm$^5$ and $S_{\Sigma}
= -22.3$ MeV$\cdot$fm$^5$ at $k_F = 1.35$ fm$^{-1}$
into various contributions.
The model is FSS. The unit is $\hbox{MeV}\cdot \hbox{fm}^5$.}
\label{table8}
\bigskip
\setlength{\tabcolsep}{4mm}
\renewcommand{\arraystretch}{1.4}
\begin{center}
\begin{tabular}{@{}c@{\extracolsep{\fill}}rrrrrr}
\hline
 &  & \multicolumn{2}{c}{$I=1/2$} & &
  \multicolumn{2}{c}{$I=3/2$}\\
 \cline{3-4} \cline{6-7}
 &  & odd & even &
 & odd & even \\ \hline
 $S_{\Lambda}$ & $LS$    & $-17.1$ & $0.6$ & & --- & --- \\
 & $LS^{(-)}$  & $12.7$  & $0.3$   & & --- & --- \\ \hline
 $S_{\Sigma}$  & $LS$    & $2.7$  & $0.1$ & & $-11.9$ & $-1.6$ \\
 & $LS^{(-)}$  & $-10.5$ & $-0.6$  & & $0.1$ & $-0.0$ \\
\hline
\end{tabular}
\end{center}
\end{table}
%

\section{Summary}

Since the spin-orbit force is the simplest
momentum-dependent short-range force in the baryon-baryon
interaction, it is sometimes discussed that
the quark substructure of baryons might play an essential
role as the microscopic origin of this very important
non-central force \cite{SU84,MO84,HE86}.
In the hyperon-nucleon ($YN$) interaction,
the spin-orbit force has very rich contents, consisting
of three different types; $LS$, $LS^{(-)}$,
and $LS^{(-)}\sigma$ \cite{FU97}.
These $LS$ forces predicted by the color-analogue
of the Fermi-Breit (FB) interaction
in the $(3q)$-$(3q)$ resonating-group method have
correct spin and flavor dependence,
which is very similar to that predicted by traditional
meson-exchange models \cite{NA93}.
As to the magnitude of these $LS$ forces, we have pointed
out \cite{SU84} that the inclusion
of the Galilean non-invariant $aLS$ term
of the FB interaction is important,
since it gives almost one-third
of the Galilean invariant $sLS$ term with the same sign.
The choice of the harmonic oscillator constant $b$ is also crucial
to obtain enough strength of the $LS$ forces.
In order to confirm that these $LS$ forces
are consistently described with the short-range repulsion,
we have proposed several unified models
of the $NN$ and $YN$ interactions \cite{NA95,FU95,FU96a,FU96b,FJ98},
in which a realistic description
of these interactions is achieved not only for the $LS$ forces
but also for many other components of the central
and non-central forces.
In these models, the short-range interaction composed of
the strongly repulsive central force and the $LS$ forces
is mainly described by the quark-exchange kernel
of the FB interaction, and the medium- and long-range interaction
composed of the attractive central force
and the long-range tensor force is described
by meson-exchange processes acting between quarks.

In this paper we have developed a formulation
of the single-particle (s.p.) spin-orbit ($\ell s$) potentials
for the nucleon and hyperons, following the idea presented
by Scheerbaum \cite{SC76}.
The quark-exchange kernel from the color-analogue
of the FB interaction is
directly employed to calculate the strength
factor $S_B$ for the s.p. $\ell s$ potentials in the
Born approximation. In the simplest treatment, $S_B$ is
concisely expressed in terms of quark parameters,
among which the parameter $\lambda=(m_s/m_{ud})$,
representing the flavor-$SU_3$ symmetry
breaking (FSB) at the quark level, plays an important role. 
Such expressions are very useful for examining
the characteristic structure of the s.p. $\ell s$ potentials.
The ratio of $S_B$ to the nucleon strength $S_N$ for the
spin-saturated $Z=N$ nuclei is found
to be $S_\Lambda/S_N \sim 1/5$, $S_\Sigma/S_N \sim 1/2$
and $S_\Xi/S_N \sim -1/4$ in the Born approximation
with the full FSB,
irrespective of various versions of our quark model.
This result is consistent with the estimation by 
Morimatsu {\em et al.} \cite{MO84},
$U_N : U_\Lambda : U_\Sigma=1 : 0.21 : 0.55$,
although they used only the
Galilean invariant $sLS$ term in the FB interaction.
This ratio is also preserved by the Galilean
non-invariant $aLS$ term in the FB interaction,
but the inclusion of this term makes the magnitude
of $S_B$ reasonable for the realistic description,
in contrast to the large value presented in \cite{MO84}.
It is interesting to note that Dover and Gal \cite{DO84} also
predicted $V_{SO}^\Lambda/V_{SO}^N
\sim 0.2$ and $V_{SO}^\Sigma/V_{SO}^N \sim 0.6$,
by using the coupling constants of the Nijmegen model F potential.
 
We have also developed a formulation to evaluate the $S_B$ factor
from the $G$-matrix solution of our quark-model potential. 
Here we first calculated $NN$, $\Lambda N$ and $\Sigma N$ $G$-matrices
in symmetric nuclear matter
by solving the Bethe-Goldstone equation for the exchange kernel
of our quark model FSS \cite{FU96a,FU96b}.
These $G$-matrices are then used to calculate $S_B$ for
spin-saturated symmetric nuclear matter,
in the same way as the calculation of the s.p. potentials.
In the limit of the zero-momentum hyperons, we have found
a fairly large reduction of $S_\Lambda$,
resulting in the ratio $S_\Lambda/S_N \sim 1/12$.
For $S_N$ and $S_\Sigma$, the effect produced by solving
the $G$-matrix equation is comparatively weak against usual
phenomenological potentials with a short-range repulsive core.
In particular, we have
found $S_N \sim -40~\hbox{MeV}\cdot\hbox{fm}^5$ both
in the Born approximation and in the $G$-matrix calculation.
This implies that the effect of the shot-range correlation is
rather moderate in the quark-model description of the short-range
repulsion.

In the hyperon s.p. $\ell s$ potentials,
the antisymmetric $LS$ ($LS^{(-)}$) force
originating from the FB spin-orbit interaction 
(both from the $sLS$ and $aLS$ pieces) plays a characteristic role.
If we neglect the FSB, the $S_\Lambda/S_N$ ratio is already around 1/3.
This is because the half of the $LS$ contribution is cancelled
by the $LS^{(-)}$ contribution. The ratio is further reduced
to 1/5 by the FSB, originating
from the strange to up-down quark-mass difference
and the reduction factor of the $LS$ operator due to the difference
of $N$ and $\Lambda$ baryon masses. The former feature of the FSB
at the quark level is a special situation of the $\Lambda$ hyperon,
which results from the structure
of its spin-flavor $SU_6$ wave function.
Finally, the short-range correlation by solving the $G$-matrix
equation further reduces the ratio
to less than 1/10.
It may be argued that the $\ell s$ potential is relevant
at the surface region in finite nuclei, where the nucleon density
is rather low. We have checked that a small deviation
of the Fermi-momentum from the value of
ordinary symmetric nuclear matter,
$k_F=1.35~\hbox{fm}^{-1}$, does not change this small ratio.
Experimental confirmation of the small s.p. $\ell s$ potentials
for the $\Lambda$ hyperon is highly desirable \cite{TA99}.

\bigskip

\appendix

\section{Isospin factors $C_\tau^I(B)$ and spatial
integrals $f_\CT^{LS}(\theta)$}

In this appendix we give explicit expressions
of the isospin factors $C_\tau^I(B)$ in
\eq{fm13} and the spatial integrals $f_\CT^{LS}(\theta)$ in \eq{fm9}.
The explicit values of $C_\tau^I(B)$ are
\begin{eqnarray}
& & C_p^I(p)=C_n^I(n)=C_p^I(\Xi^0)
=C_n^I(\Xi^-)=\delta_{I,1}\ \ ,\nonumber \\
& & C_p^I(n)=C_n^I(p)=C_p^I(\Xi^-)
=C_n^I(\Xi^0)={1 \over 2} \qquad \hbox{for~both}~I=0
~\hbox{and}~I=1\ \ ,\nonumber \\
& & C_p^{1 \over 2}(\Lambda)
=C_n^{1 \over 2}(\Lambda)=1\ \ ,\nonumber \\
& & C_p^I(\Sigma^+)=C_n^I(\Sigma^-)
=\delta_{I, {3 \over 2}}\ \ ,\nonumber \\
& & C_p^I(\Sigma^-)=C_n^I(\Sigma^+)
=\delta_{I, {1 \over 2}}~{2 \over 3}
+\delta_{I, {3 \over 2}}~{1 \over 3}\ \ ,\nonumber \\
& & C_p^I(\Sigma^0)=C_n^I(\Sigma^0)
=\delta_{I, {1 \over 2}}~{1 \over 3}
+\delta_{I, {3 \over 2}}~{2 \over 3}\ \ ,
\label{a1}
\end{eqnarray}
and a sum rule
\begin{equation}
\sum_\tau C_\tau^I(B)={2I+1 \over 2I_B+1}
\label{a2}
\end{equation}
is satisfied for each $B$.
The spatial functions $f_\CT^{LS}(\theta)$ are given
by\footnote{See Appendix B of \protect\cite{LS99}.} 
\begin{eqnarray}
& & f^{LS}_\CT(\theta)=(-2\pi)
\,\alpha_S x^3 m_{ud}c^2~b^5 \nonumber \\
& & \times \left\{
\begin{array}{c}
\left({12 \over 11}\right)^{{3 \over 2}}
\exp \left\{-{2 \over 11} b^2 \left[{4 \over 3}(\bq^2+\bk^2)
-\bk \cdot \bq \right] \right\}
~\widetilde{h}_1 \left({1 \over \sqrt{11}}b |\bq+\bk| \right) \\
\left({12 \over 11}\right)^{{3 \over 2}}
\exp \left\{-{2 \over 11} b^2 \left[{4 \over 3}(\bq^2+\bk^2)
+\bk \cdot \bq \right] \right\}
~\widetilde{h}_1 \left({1 \over \sqrt{11}}b |\bq-\bk| \right) \\
\left({3 \over 4}\right)^{{3 \over 2}}
\exp \left\{ -{1 \over 3} b^2 \left(\bq^2+{1 \over 4} \bk^2\right)
\right\}
~\widetilde{h}_1 \left({1 \over 2} b |\bk| \right) \\
\exp \left\{ -{1 \over 3} b^2 \bk^2 \right\}
~\widetilde{h}_1 \left({1 \over \sqrt{3}} b |\bq| \right) \\
\end{array}
\right. \nonumber \\
& & \ \ \hspace{70mm} \hbox{for} \quad \CT= \left\{ \begin{array}{c}
S \\
S^\prime \\
D_+ \\
D_- \\
\end{array}
\ \ ,\right.
\label{a3}
\end{eqnarray}
where $\bk=\bq_f-\bq_i$, $\bq=(1/2)(\bq_f+\bq_i)$,
$\cos \theta=\widehat{\bq}_f \cdot \widehat{\bq}_i$,
and $\widetilde{h}_1(x)$ is defined as
\begin{equation}
\widetilde{h}_1(x) = 3 e^{-x^2}\int^1_0
e^{x^2t^2} t^2 d t = 1+3 \sum_{n=1}^\infty
(-1)^n {(2x^2)^n \over (2n+3)!!} \ \ .
\label{a4}
\end{equation}
Here $\widetilde{h}_1(x)$ is normalized as $\widetilde{h}_1(0)=1$.
If we set $\bq=0$ in \eq{a3}, it is simplified to
\begin{eqnarray}
& & \overline{f^{LS}_\CT(\theta)} \sim
f^{LS}_\CT(\theta) \vert_{\bq=0}=(-2\pi)
\,\alpha_S x^3 m_{ud}c^2~b^5 \nonumber \\
& & \times \left\{
\begin{array}{c}
\left({12 \over 11}\right)^{{3 \over 2}}
\exp \left\{ -{8 \over 33}(bk)^2 \right\}
~\widetilde{h}_1 \left({1 \over \sqrt{11}}bk\right) \\
\left({3 \over 4}\right)^{{3 \over 2}}
\exp \left\{ -{1 \over 12}(bk)^2 \right\}
~\widetilde{h}_1 \left({1 \over 2}bk\right) \\
\exp \left\{ -{1 \over 3}(bk)^2 \right\} \\
\end{array}
\right. \quad \hbox{for} \quad \CT= \left\{ \begin{array}{c}
S,~S^\prime \\
D_+ \\
D_- \\
\end{array}
\right.,
\label{a5}
\end{eqnarray}
where we assume $k=\overline{q}\sim (2\pi/4t)
\sim 0.7~\hbox{fm}^{-1}$.
The analytic expression of $S_B$ in \eq{re2} is easily derived,
if we further set $\overline{q}=0$.

\end{document}